\documentclass[10pt,conference]{IEEEtran}
\usepackage{epsfig,rotating,setspace,latexsym,amsmath,epsf,amssymb,bm,theorem}
\usepackage{cite}

\newtheorem{definition}{Definition}

\IEEEoverridecommandlockouts

\begin{document}

\title{Polar Coding for the General Wiretap Channel\thanks{This work was supported by NSF Grants CNS 13-14733, CCF 14-22111 and CCF 14-22129.}}

\author{\IEEEauthorblockN{Yi-Peng Wei \qquad Sennur Ulukus}
\IEEEauthorblockA{Department of Electrical and Computer Engineering\\
University of Maryland College Park, MD 20742\\
{\it ypwei@umd.edu \qquad ulukus@umd.edu} }}

\maketitle

\begin{abstract}
Information-theoretic work for wiretap channels is mostly based on random coding schemes. Designing practical coding schemes to achieve information-theoretic security is an important problem. By applying the two recently developed techniques for polar codes, we propose a polar coding scheme to achieve the secrecy capacity of the general wiretap channel.
\end{abstract}

\section{Introduction}

The wiretap channel was first introduced by Wyner \cite{Wyner_WTC_75}, in which a legitimate transmitter (Alice) wishes to send messages to a legitimate receiver (Bob) secretly in the presence of an eavesdropper (Eve). Wyner \cite{Wyner_WTC_75} characterized the capacity equivocation region for the degraded wiretap channel, in which the received signal at Eve is a degraded version of the received signal at Bob. Later, Csisz\'{a}r and K\"{o}rner \cite{Csiszar_IT_78} characterized the capacity equivocation region for general, not necessarily degraded, wiretap channels. These works are based on information-theoretic random coding schemes.

Polar coding, invented by Ar{\i}kan \cite{Arikan_IT_09}, is the first code that provably achieves the capacity of the binary-input discrete symmetric output channels (B-DMC). The idea of polar coding has been extended to
lossless source coding \cite{Arikan_ISIT_10}, lossy source coding \cite{Korada_IT_10}, and to multi-user scenarios, such as, multiple access channel \cite{Sasoglu_IT_13, Abbe_IT_12, Onay_ISIT_13}, broadcast channel \cite{Gastpar_arXiv_13, Marco_arXiv_14}, interference channel \cite{Wang_arXiv_14}, and Slepian-Wolf coding problem \cite{Arikan_ISIT_12}.

On a B-DMC, polarization results in two kinds of sub-channels \cite{Arikan_IT_09}. The first kind is good sub-channels. The capacity for these sub-channels approaches $1$ bit per channel use. The second kind is bad sub-channels. The channel output for these sub-channels is independent of the channel input; therefore the capacity for these sub-channels approaches $0$. In particular, if a B-DMC A is degraded with respect to a B-DMC B, then the good sub-channels of A must be a subset of the good sub-channels of B \cite{Korada_thesis_09}. We call this the \emph{subset property}.

Polar coding schemes for \emph{degraded} wiretap channels with \emph{symmetric} main and eavesdropper channels are developed using the subset property in \cite{Vardy_IT_11, Skoglund_Comm_Letter_10, El_Gamal_PIMRC_10, Shamai_ITW_10}. For degraded wiretap channels, the good sub-channels of Eve is a subset of the good sub-channels of Bob. The polar coding scheme is designed to transmit the confusion messages (random bits) on the sub-channels simultaneously good for Bob and Eve, and to transmit the secret messages on the sub-channels only good for Bob. However, for non-degraded wiretap channels, the subset property no longer holds \cite{Hassani_Allerton_09, Vardy_IT_13,  Sutter_arXiv_14, Hassani_arXiv_13, Sasoglu_arXiv_13}, i.e., the good sub-channels of Eve is not necessary a subset of the good sub-channels of Bob. Moreover, the secrecy capacity achieving input distribution is not necessarily a uniform distribution. Therefore, the polar coding schemes in \cite{Vardy_IT_11, Skoglund_Comm_Letter_10, El_Gamal_PIMRC_10, Shamai_ITW_10} cannot directly extend to the non-degraded wiretap channel.

By applying the two recently developed techniques for polar codes, we can achieve the secrecy capacity of the general wiretap channel. The first is universal polar codes \cite{Hassani_arXiv_13, Sasoglu_arXiv_13}. Universal polar coding allows us to align the good sub-channels of Bob and Eve together. Therefore, we can artificially construct the subset property for the non-degraded wiretap channel. Then, Alice transmits the random bits on the sub-channels simultaneously good for Bob and Eve, and the secret message on the sub-channels only good for Bob. The second is polar coding for asymmetric models \cite{Honda_IT_13}, which allows us to deal with the non-uniform input distribution. Different from B-DMC, polarization for asymmetric channel results in three different kinds of sub-channels.

Another polar coding scheme for the general wiretap channel is provided in \cite{sutter_Arxiv_13}, which uses a concatenated code consisting of two polar codes. The inner layer ensures that the transmitted message can be reliably decoded by Bob, and the outer layer guarantees that the message is kept secret from Eve. Our work jointly handles these two goals in one shot. Hence, the decoding error probability of our scheme is approximately $O(2^{ -n^{1/2}} )$, whereas it is $O(\sqrt{n}2^{-n^{1/4}})$ in \cite{sutter_Arxiv_13}. Moreover, for practical code construction, there is still no efficient way to characterize the outer index set \cite[Sec~III. C.]{sutter_Arxiv_13}, while our coding scheme can be efficiently constructed by \cite{Vardy_IT_13}.

\section{Wiretap Channel Model} \label{Sec_model}

A wiretap channel consists of a legitimate transmitter (Alice) who wishes to send messages to a legitimate receiver (Bob) secretly in the presence of an the eavesdropper (Eve). The channel between Alice and Bob is called the main channel, and the channel between Alice and Eve is called the eavesdropper channel. Let $X$ denote the single-letter input to the main and eavesdropper channels. Let $Y$ and $Z$ denote the corresponding single-letter outputs of the main and the eavesdropper channels, respectively. $W$ represents the message to be sent to Bob and kept secret from Eve with $W \in \mathcal{W}=\{1,\cdots, 2^{nR}\}$. Let $P_e=\mathrm{Pr(\hat{W} \neq W)}$ denote the probability of error for Bob's decoding.

The equivocation rate is given by $\frac{1}{n}H(W|Z^n)$, which reflects the uncertainty of the message given eavesdropper's channel observation. A rate-equivocation pair $(R,R_e)$ is achievable if as $n \rightarrow \infty$, $P_e \rightarrow 0$  and $\lim_{n\rightarrow \infty} \frac{1}{n} H(W|Z^n) \geq R_e$. Perfect (weak) secrecy is achieved if $R=R_e$ \cite{Csiszar_IT_78}. Therefore, perfect secrecy is achieved if $\frac{1}{n} I(W;Z^n) \rightarrow 0$, and the \emph{secrecy capacity} $C_s$ is the highest achievable perfect secrecy rate $R$, which is also the highest possible equivocation rate \cite{Csiszar_IT_78}. Csisz\'{a}r and K\"{o}rner characterized the secrecy capacity for the general wiretap channel, which is \cite{Csiszar_IT_78}
\begin{equation} \label{C_s}
C_s= \max_{V \rightarrow X \rightarrow Y,Z } I(V;Y)-I(V;Z).
\end{equation}

In the following, we assume that we already know the optimal input distribution \cite{Omur_IT_13}, i.e., we know the optimal $V$, $X$ that achieve $C_s$. Although we focus on developing a coding scheme for binary inputs below, there is no difficulty to extend the work to $q$-ary inputs \cite{Sasoglu_ITW_09, Mori_ISIT_10, sasoglu_ISIT_12, Barg_IT_13}.

\section{Polar Codes } \label{Sec_Polar_code}

\subsection{Polar Codes for Asymmetric Channels} \label{Sec_polar_code_asymmetric}

Let $P_{TV}$ be the joint distribution of a pair of random variables $(T,V)$, where $T$ is a binary random variable and $V$ is any finite alphabet random variable. Let us define the Bhattacharyya parameter as follows
\begin{equation}\label{Z_parameter}
Z(T|V)=2 \sum \limits_{v} P_V(v) \sqrt{P_{T|V}(0|v)P_{T|V}(1|v)}.
\end{equation}
Let $U^n=X^nG_n$, where $X^n$ denotes $n$ independent copies of the random variable $X$ with $X \sim P_X$, and
$G_n=G^{\otimes k}$ where $G=\begin{bmatrix}
       1 & 0 \\
       1 & 1
\end{bmatrix}$
and $\otimes$ denotes the Kronecker product of matrices for $n=2^k$. \cite{Arikan_ISIT_10} shows as $n\to\infty$, $U_i$ is almost independent of $U^{i-1}$ and uniformly distributed, or otherwise $U_i$ is almost determined by $U^{i-1}$. Therefore, $[n]$, the index set $\{1,2,\ldots,n\}$, is almost polarized into two sets $\mathcal{H}_X$ and $\mathcal{L}_X$:
\begin{align}
\mathcal{H}_X &= \{ i\in[n]:Z(U_i|U^{i-1}) \geq 1- \delta_n \} \notag \\
\mathcal{L}_X &= \{ i\in[n]:Z(U_i|U^{i-1}) \leq \delta_n \}, \label{H_X_L_X_def}
\end{align}
where $\delta_n=2^{-n^\beta}$ and $\beta \in (0, 1/2)$. Moreover,
\begin{align}
\lim_{n\to\infty} \frac{1}{n} |\mathcal{H}_X|&=H(X)    \notag \\
\lim_{n\to\infty} \frac{1}{n} |\mathcal{L}_X|&=1-H(X). \label{H_X_L_X_ratio}
\end{align}

Let $P$ be a discrete memoryless channel with a binary input $X$ and finite alphabet output $Y$. Here, $P$ does not have to be a symmetric channel. Fix a distribution $P_X$ for $X$. \cite{Honda_IT_13} generalizes the above argument to achieve a rate close to $I(X;Y)$. Consider two subsets of $[n]$, $\mathcal{H}_{X|Y}$ and $\mathcal{L}_{X|Y}$, defined as follows
\begin{align}
\mathcal{H}_{X|Y} &= \{ i\in[n]:Z(U_i|U^{i-1}, Y^n) \geq 1- \delta_n  \} \notag \\
\mathcal{L}_{X|Y} &= \{ i\in[n]:Z(U_i|U^{i-1}, Y^n) \leq \delta_n  \}, \label{H_X|Y_L_X|Y_def}
\end{align}
similar to \eqref{H_X_L_X_ratio}, we have
\begin{align}
\lim_{n\to\infty} \frac{1}{n} |\mathcal{H}_{X|Y}|&=H(X|Y)    \notag \\
\lim_{n\to\infty} \frac{1}{n} |\mathcal{L}_{X|Y}|&=1-H(X|Y). \label{H_X|Y_L_X|Y_ratio}
\end{align}

With \eqref{H_X_L_X_def} and \eqref{H_X|Y_L_X|Y_def}, we define the following three sets
\begin{align}
\mathcal{I}    &= \mathcal{H}_X \cap \mathcal{L}_{X|Y}      \label{I_def}    \\
\mathcal{F}_r  &= \mathcal{H}_X \cap \mathcal{L}_{X|Y}^\mathrm{c} \label{Fr_def}     \\
\mathcal{F}_d  &= \mathcal{H}_X^\mathrm{c}. \label{Fd_def}
\end{align}
In the following, we call the set $\mathcal{I}$ the \emph{information set}, and sets $\mathcal{F}_r$ and $\mathcal{F}_d$ the \emph{frozen set}. Although we call them the \emph{frozen set}, $\mathcal{F}_r$ and $\mathcal{F}_d$ have different operational meanings which will be illustrated below. Note that for the symmetric channel capacity achieving code design, $\mathcal{F}_d$ is an empty set \cite{Arikan_IT_09}.

To achieve rate $I(X;Y)$ for channel $P$, let us consider the following coding scheme. First, the encoder transmits the information bits in the index set $\mathcal{I}$. For $i \in \mathcal{I}$ in \eqref{I_def}, since $i \in \mathcal{H}_X$, $U_i$ is almost independent of $U^{i-1}$ and uniformly distributed. Therefore, the encoder can freely assign values to $U_\mathcal{I}$, where $U_\mathcal{I}$ denotes a sub-vector $\{U_i\}_{i\in\mathcal{I}}$. Moreover, since $i \in \mathcal{L}_{X|Y}$, $U_i$ is almost determined by $U^{i-1}$ and $Y^n$, which means that given the channel output $Y^n$, $U_i$ is decoded in a successive manner.

Second, for $i \in \mathcal{F}_r$ in \eqref{Fr_def}, $U_i$ is almost independent of $U^{i-1}$ and uniformly distributed, and given the channel output $Y^n$, $U_i$ cannot be reliably decoded. The encoder transmits $U_{\mathcal{F}_r}$ with a uniformly random sequence and the randomness is shared between the transmitter and the receiver.

Last, for $i \in \mathcal{F}_d$ in \eqref{Fd_def}, $U_i$ is almost determined by $U^{i-1}$. The values of $U_{ \mathcal{F}_d}$ are computed in successive order through the following randomized map
\begin{equation} \label{u_i_F_d}
u_i = \arg \max_{u \in \{0,1\}} P_{U_i|U^{i-1}}(u|u^{i-1}).
\end{equation}

By \eqref{H_X_L_X_ratio} and \eqref{H_X|Y_L_X|Y_ratio}, it is easy to verify that
\begin{equation} \label{I(X;Y)}
\lim_{n\to\infty} \frac{1}{n} |\mathcal{I}|=I(X;Y).
\end{equation}
Moreover, by applying successive cancellation decoder, the block error probability $P_e$ can be upper bounded by
\begin{equation} \label{Pe_bound}
P_e \leq \sum_{i \in \mathcal{I} } Z(U_i|U^{i-1}, Y^n)=O(2^{-n^\beta})
\end{equation}
for any $\beta \in (0, 1/2)$, with complexity $O(n \log n)$. Therefore, the rate $I(X;Y)$ is achieved.

\subsection{Universal Polar Coding}
Consider two B-DMCs $P: X \rightarrow Y$ and $Q: X \rightarrow Z$, and assume that these two channels have identical capacities, i.e., $C(P)=C(Q)$. Let $U^n=X^nG_n$, and denote $\mathcal{P}$ and $\mathcal{Q}$ as
the information set defined in \eqref{I_def}, i.e.,
\begin{align}
\mathcal{P} &= \{ i \in [n]: Z(U_i|U^{i-1}, Y^n)  \leq \delta_n  \}  \notag \\
\mathcal{Q} &= \{ i \in [n]: Z(U_i|U^{i-1}, Z^n)  \leq \delta_n  \}, \notag
\end{align}
where $\delta_n=2^{-n^\beta}$ and $\beta \in (0, 1/2)$. Since we assume $C(P)=C(Q)$, we also have $|\mathcal{P}|=|\mathcal{Q}|$.

In general, the differences $\mathcal{P} \setminus \mathcal{Q}$ and $\mathcal{Q} \setminus \mathcal{P}$ are not empty sets \cite{Hassani_Allerton_09, Vardy_IT_13,  Sutter_arXiv_14}; therefore, it is not straightforward to apply standard polar coding to achieve the capacity of the compound channel consisting of $P$ and $Q$. \cite{Hassani_arXiv_13} proposes a method, called \emph{chaining construction}, to solve this problem.

\begin{definition} (Chaining construction \cite{Hassani_arXiv_13}) \label{chain}
Let $m \geq 2$. The $m$-chain of $\mathcal{P}$ and $\mathcal{Q}$ is a code of length $mn$ that consists of $m$ polar blocks of length $n$. In each of the $m$ blocks, the set $\mathcal{P} \cap \mathcal{Q}$ is set to be an information set. In the $i$th block, $1 \leq i <m$, the set $\mathcal{P} \setminus \mathcal{Q}$ is also set to be an information set. Moreover, the set $\mathcal{P} \setminus \mathcal{Q}$ in the $i$th block is chained to the set $\mathcal{Q} \setminus \mathcal{P}$ in the $(i+1)$th block in the sense that the information is repeated in these two sets. All other indices are frozen. Therefore, in each block, the set $(\mathcal{P} \cup \mathcal{Q})^c$ is frozen, and the set $\mathcal{Q} \setminus \mathcal{P}$ in the $1$st block and the set $\mathcal{P} \setminus \mathcal{Q}$ in the $m$th block are frozen, too. Note that $(\cdot)^c$ denotes the complement of a set. The rate of the chaining construction is
\begin{align} \label{compound_rate}
\frac{|\mathcal{P} \cap \mathcal{Q}|+\frac{m-1}{m}|\mathcal{P} \setminus \mathcal{Q}|}{n}.
\end{align}
\end{definition}

Next, we discuss the decoding procedure for the compound channel consisting of $P$ and $Q$. If the channel $P$ is used, then we decode from the first block. On the other hand, if the channel $Q$ is used, then we decode from the last block.

First, suppose that channel $P$ is used and a code of length $mn$ has been received. For this case, we decode from the first block. In the $1$st block, we put all the information bits in the set $\mathcal{P}$, thus the decoder can decode correctly. For the $2$nd block, through chaining construction, the set $\mathcal{P} \setminus \mathcal{Q}$ in the $1$st block is chained to the set $\mathcal{Q} \setminus \mathcal{P}$ in the $2$nd block, and the set $(\mathcal{P} \cup \mathcal{Q})^c$ is frozen. Equivalently, the decoder only needs to decode the bits in the set $\mathcal{P}$, which can be correctly decoded. The same procedure holds until the $(m-1)$th block. For the $m$th block, the information bits are only put in the set $\mathcal{P} \cap \mathcal{Q}$, and the remaining part has been determined. Hence, information bits can be reliably decoded. The main rate loss for the chaining construction comes from the last block.

Second, consider the case that the channel $Q$ is used. In this case, we decode from the last block. In the $m$th block, since the information bits are put in the set $\mathcal{Q}$, reliable decoding is guaranteed. For the $(m-1)$th block, due to the chaining process, the set $\mathcal{Q} \setminus \mathcal{P}$ in the $m$th block is chained to the set $\mathcal{P} \setminus \mathcal{Q}$ in the $(m-1)$th block, and note that the set $(\mathcal{P} \cup \mathcal{Q})^c$ is frozen. The decoder only needs to decode the information bits in the set $\mathcal{Q}$, thus correct decoding is ensured. This procedure is applied until the $2$nd block. For the $1$st block, information bits which have not been determined fall in the set $\mathcal{P} \cap \mathcal{Q}$, thus the decoder can decode them correctly.

In summary, for a fixed $m$, if we let $n \rightarrow \infty$, we can achieve the rate in \eqref{compound_rate} with arbitrary small error probability, which also means that the rate $C(P)-\frac{1}{m}\frac{|\mathcal{P} \setminus \mathcal{Q}|}{n}$ can be achieved. Additionally, if we let $m\rightarrow \infty$, then the rate $C(P)$, which is the capacity of the compound channel consisting of channels $P$ and $Q$, can be achieved.

\section{Polar coding for the general wiretap channel} \label{Sec_Coding_Scheme}

Assume now that we know the optimal distributions to achieve the secrecy capacity $C_s$ in \eqref{C_s}, i.e., we know the optimal $V$ and $X$. For illustration, we consider the case of a binary input channel, i.e., $|\mathcal{X}|=2$. The cardinality bound for channel prefixing $V$, is $|\mathcal{V}|\leq2$.

\subsection{The Scheme} \label{Sec_scheme_design}

Let $U^n=V^nG_n$. Consider the following sets:
\begin{align}
\mathcal{H}_V &= \{ i\in[n]:Z(U_i|U^{i-1}) \geq 1- \delta_n \} \notag \\
\mathcal{L}_V &= \{ i\in[n]:Z(U_i|U^{i-1}) \leq \delta_n \}, \label{H_V_L_V_def} \\
\mathcal{H}_{V|Y} &= \{ i\in[n]:Z(U_i|U^{i-1}, Y^n) \geq 1- \delta_n  \} \notag \\
\mathcal{L}_{V|Y} &= \{ i\in[n]:Z(U_i|U^{i-1}, Y^n) \leq \delta_n  \}, \label{H_V|Y_L_V|Y_def} \\
\mathcal{H}_{V|Z} &= \{ i\in[n]:Z(U_i|U^{i-1}, Z^n) \geq 1- \delta_n  \} \notag \\
\mathcal{L}_{V|Z} &= \{ i\in[n]:Z(U_i|U^{i-1}, Z^n) \leq \delta_n  \}, \label{H_V|Z_L_V|Z_def}
\end{align}
where $\delta_n=2^{-n^\beta}$ and $\beta \in (0, 1/2)$.

The set $[n]$ can be partitioned into the following four sets:
\begin{align}
G_{Y\wedge Z}    &= \mathcal{H}_V \cap \mathcal{L}_{V|Y} \cap \mathcal{L}_{V|Z}, \\
G_{Y\setminus Z} &= \mathcal{H}_V \cap \mathcal{L}_{V|Y} \cap \mathcal{L}_{V|Z}^c, \\
G_{Z\setminus Y} &= \mathcal{H}_V \cap \mathcal{L}_{V|Y}^c \cap \mathcal{L}_{V|Z}, \\
B_{Y\wedge Z}    &= \mathcal{H}_V^c \cup (\mathcal{L}_{V|Y}^c \cap \mathcal{L}_{V|Z}^c).
\end{align}
From a successive decoding point of view, the sub-channels corresponding to the set $G_{Y\wedge Z}$ are simultaneously good for Bob and Eve. The sub-channels in the set $G_{Y\setminus Z}$ are good for Bob but bad for Eve. On the other hand, the sub-channels in the set $G_{Z\setminus Y}$ are good for Eve but bad for Bob. Last, the sub-channels in the set $B_{Y\wedge Z}$ are bad for both Bob and Eve.

Similar to  \eqref{I_def}--\eqref{Fd_def}, we have:
\begin{align}
\mathcal{I}_Y &= \mathcal{H}_V \cap \mathcal{L}_{V|Y}, \notag \\
\mathcal{I}_Z &= \mathcal{H}_V \cap \mathcal{L}_{V|Z}, \notag \\
\mathcal{F}_r^Y &= \mathcal{H}_V \cap \mathcal{L}_{V|Y}^c,  \notag \\
\mathcal{F}_r^Z &= \mathcal{H}_V \cap \mathcal{L}_{V|Z}^c,  \notag \\
\mathcal{F}_d   &= \mathcal{H}_V^c.
\end{align}
By \eqref{I(X;Y)}, we have
\begin{align} \label{Bob_Eve_rate}
\lim_{n\to\infty} \frac{1}{n} |\mathcal{I}_Y|&=I(V;Y),  \notag \\
\lim_{n\to\infty} \frac{1}{n} |\mathcal{I}_Z|&=I(V;Z).
\end{align}

For the \emph{symmetric} and \emph{degraded} wiretap channel \cite{Vardy_IT_11, Skoglund_Comm_Letter_10, El_Gamal_PIMRC_10, Shamai_ITW_10}, $G_{Z\setminus Y}$ is an empty set, since the degraded property of the channel causes $\mathcal{I}_Z \subset  \mathcal{I}_Y$ \cite{Korada_thesis_09}. However, for the general wiretap channel, $G_{Z\setminus Y}$ is no longer an empty set, and $|G_{Z\setminus Y}|$ cannot be negligible \cite{Hassani_Allerton_09, Vardy_IT_13,  Sutter_arXiv_14}.

Here, we consider the positive secrecy capacity case, thus we have $|G_{Y\setminus Z}| > |G_{Z\setminus Y}|$. Choose a set, $C_{Y\setminus Z}$, such that $C_{Y\setminus Z} \subset G_{Y\setminus Z} $ and $|C_{Y\setminus Z}|=|G_{Z\setminus Y}|$. Define the set $S$ as:
\begin{equation}
S=G_{Y\setminus Z} \setminus C_{Y\setminus Z} . \notag
\end{equation}
From \eqref{Bob_Eve_rate}, we have
\begin{equation} \label{R_s}
\lim_{n\to\infty} \frac{1}{n} |S|=I(V;Y)-I(V;Z).
\end{equation}

\begin{figure}
\centering
\epsfig{file=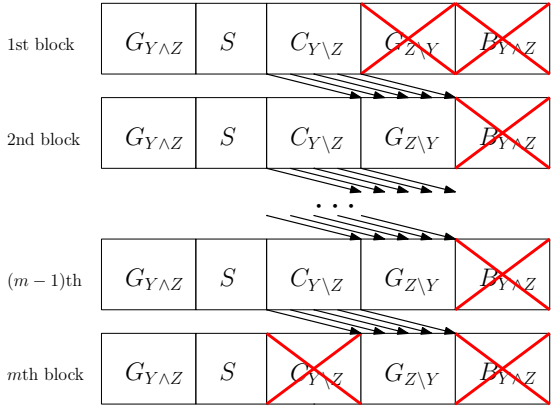,width=0.40\textwidth}
\caption{Chaining construction.}
\label{Fig_chain}
\vspace*{-0.4cm}
\end{figure}

We construct the code as follows. Consider an \emph{m-chain} polar code in Definition~\ref{chain}. For $1 \leq i <m$, the set $C_{Y \setminus Z}$ in the $i$th block is chained to $G_{Z \setminus Y}$ in the $(i+1)$th block as in Fig.~\ref{Fig_chain}. For each of the $m$ blocks, the set $B_{Y \wedge Z}$ is set to be frozen. Moreover, the set $G_{Z \setminus Y}$ in the $1$st block is set to be frozen in the sense that $G_{Z \setminus Y} \subseteq \mathcal{F}_r^Y$, and the set $C_{Y \setminus Z}$ in the $m$th block is also set to be frozen in the sense that $C_{Y \setminus Z }\subseteq \mathcal{F}_r^Z$. In Fig.~\ref{Fig_chain}, we use a red cross to denote a frozen set.

We put the secret information bits in the set $S$ in each block. Therefore, the set $S$ is used for secret message transmission. For blocks $1 \leq i <m$, we put uniformly distributed random bits to $C_{Y \setminus Z}$
to serve as the confusion messages. Through the chaining construction, the confusion messages are also chained to the set $G_{Z \setminus Y}$ in block $1 <  i \leq m$. Moreover, the set $G_{Y \wedge Z}$ in each block are also filled with random bits to serve as confusion message. For the frozen sets, if the index belongs to $\mathcal{F}_r^Y$ or $\mathcal{F}_r^Z$, then we put uniformly distributed random bits and share the randomness with the decoder (Bob and Eve). Last, if the index belongs to $\mathcal{F}_d$, then we determine the value according to the randomized map defined in \eqref{u_i_F_d}. We summarize the encoding procedure as follows.

\vspace*{0.2cm}
\noindent \textbf{Encoding procedure:}

\noindent For each block, the secret information bits are put in $U_S$, and determine the bits in $U_{ {\mathcal{F}_d}}$ by \eqref{u_i_F_d}.

\noindent For the $1$st block,
\begin{enumerate}
  \item Put uniformly distributed random bits to $U_{G_{Y \wedge Z} \cup C_{Y \setminus Z }}$.
  \item Put uniformly distributed random bits to $U_{ \mathcal{F}_r^Y}$, and share the randomness with the decoder.
\end{enumerate}
For the $j$th block, $2 \leq j< m$,
\begin{enumerate}
  \item Put uniformly distributed random bits to $U_{G_{Y \wedge Z} \cup C_{Y \setminus Z }}$.
  \item Chaining construction: repeat the bits in $C_{Y \setminus Z}$ of the $(j-1)$th block to the bits in $U_{ G_{Z \setminus Y}}$.
  \item Put uniformly distributed random bits to $U_{\mathcal{F}_r^Y \cap \mathcal{F}_r^Z}$, and share the randomness with the decoder.
\end{enumerate}
For the $m$th block,
\begin{enumerate}
  \item Put uniformly distributed random bits to $U_{G_{Y \wedge Z}}$.
  \item Chaining construction: repeat the bits in $C_{Y \setminus Z}$ of the $(m-1)$th block to the bits in
  $U_{ G_{Z \setminus Y}}$.
  \item Put uniformly distributed random bits to $U_{ \mathcal{F}_r^Z}$, and share the randomness with the decoder.
\end{enumerate}

Note that in the chaining construction we require the bits in $U_{G_{Z\setminus Y}}$ equal the bits in $U_{C_{Y\setminus Z}}$. Since we fill uniformly distributed random bits to $U_{C_{Y\setminus Z}}$, we simultaneously fill random bits to $U_{G_{Z\setminus Y}}$. Due to the fact that $G_{Z\setminus Y} \cap \mathcal{F}_d = \emptyset $, we can freely choose bits in this set.

\vspace*{0.2cm}
\noindent \textbf{Decoding procedure:}

\noindent Bob decodes from the $1$st block. In each block, if $i \in \mathcal{F}_d$, then $\hat{u}_i=\arg\max_{u \in \{0,1\}} P_{U_i|U^{i-1}}(u|\hat{u}^{i-1})$. For the $1$st block,
\begin{equation}
\hat{u}_i=
\left\{
  \begin{array}{l}
    u_i \\
    \hbox{if $i \in \mathcal{F}_r^Y$, } \\
    \arg\max_{u \in \{0,1\}} P_{U_i|U^{i-1}, Y^n}(u|\hat{u}^{i-1}, y^n) \\
    \hbox{if   $i \in
                         G_{Y\wedge Z} \cup
                         C_{Y\setminus Z} \cup
                         S $. }  \\
  \end{array}
\right.
\notag
\end{equation}
For the $j$th block, $2 \leq j< m$,
\begin{equation}
\hat{u}_i=
\left\{
  \begin{array}{l}
    u_i \\
    \hbox{if $i \in \mathcal{F}_r^Y \cap \mathcal{F}_r^Z $,} \\
    \arg\max_{u \in \{0,1\}} P_{U_i|U^{i-1}, Y^n}(u|\hat{u}^{i-1}, y^n) \\
    \hbox{if   $i \in  G_{Y \wedge Z} \cup C_{Y \setminus Z} \cup S $,}  \\
    \hat{u}_{i'} \text{ in the $(j-1)$th block, where } i' \in
                          C_{Y \setminus Z} \\
    \hbox{if $i \in G_{Z\setminus Y}$.} \\
  \end{array}
\right.
\notag
\end{equation}
For the $m$th block,
\begin{equation}
\hat{u}_i=
\left\{
  \begin{array}{l}
    u_i \\
    \hbox{if $i \in \mathcal{F}_r^Z $, } \\
    \arg\max_{u \in \{0,1\}} P_{U_i|U^{i-1}, Y^n}(u|\hat{u}^{i-1}, y^n) \\
    \hbox{if   $i \in    G_{Y \wedge Z} \cup S $, }  \\
    \hat{u}_{i'} \text{ in the $(m-1)$th block, where } i' \in
                          C_{Y \setminus Z} \\
   \hbox{if $i \in G_{Z\setminus Y}$.} \\
  \end{array}
\right.
\notag
\end{equation}

\subsection{Reliability}

From \eqref{R_s}, we know as $n \rightarrow \infty$, our coding scheme can achieve the secrecy rate in \eqref{C_s}. Moreover, when Bob applies the decoding procedure described in Sec.~\ref{Sec_scheme_design},
according to \eqref{Pe_bound}, the block error probability of the whole \emph{m-chain} block can be upper bounded by
\begin{align}
P_e \leq & (m-1) \sum _{i\in C_{Y \setminus Z}}
               Z(U_i|U^{i-1}, Y^n) \notag  \\
       &+  m    \sum _{i\in G_{Y \wedge Z} \cup S }
               Z(U_i|U^{i-1}, Y^n) \notag =  O(2^{-n^\beta})
\notag
\end{align}
for any $\beta \in (0, 1/2)$ with complexity $O(n \log n)$. Therefore, the secrecy rate in \eqref{C_s} can be achieved reliably.

\subsection{Equivocation Calculation}

We first introduce necessary notation for the calculation of the equivocation rate. In the encoding process, we consider $m$ blocks each with block length $n$. Let $Z^{mn}$ denote what Eve receives. For each block, we perform $U^n=V^n G_n$, therefore, for the total of $m$ blocks, we have $V^{mn}$ and $U^{mn}$.

Let $W_s$ denote the secret message, and $\tilde{W}_s$ denote the confusion message. Let the subscript $i$ of a set denote the set in the $i$th block. For example, $S_i$ denotes the set $S$ in the $i$th block, and $G_{Y \wedge Z j}$ denotes the set $G_{Y \wedge Z} $ in the $j$th block. Since secret message is put in $S_i$, $1\leq i \leq m$, we have $W_s=\cup_{1\leq i \leq m} U_{S_i}$. Also, the confusion message is put in $G_{Y \wedge Z i} , ~1\leq i \leq m $ and $C_{Y \setminus Z j}, ~1\leq j < m $. Therefore, we have $\tilde{W}_s =  \cup_{1\leq i \leq m, 1\leq j < m} U_{G_{Y \wedge Z i} } U_{C_{Y \setminus Z j}}$.

We can calculate the equivocation rate as follows:
\begin{align}
 &H(W_s|Z^{mn})  \notag\\
 &= H(W_s, \tilde{W}_s|Z^{mn})-H(\tilde{W}_s|W_s, Z^{mn})  \label{exp_a}\\
 &= H(W_s, \tilde{W}_s)-I(W_s,\tilde{W}_s;Z^{mn}) -H(\tilde{W}_s|W_s, Z^{mn}) \label{exp_b} \\
 &\geq H(W_s, \tilde{W}_s)-I(V^{mn};Z^{mn})-H(\tilde{W}_s|W_s, Z^{mn})  \label{exp_c}\\
 &= H(W_s)+H(\tilde{W}_s)-I(V^{mn};Z^{mn})-H(\tilde{W}_s|W_s, Z^{mn}) \label{exp_d}
\end{align}
which is equivalent to
\begin{align} \label{exp_e}
\frac{1}{mn} I(W_s; Z^{mn}) \leq & \frac{1}{mn}I(V^{mn};Z^{mn})+  \notag \\
                                 & \frac{1}{mn} H(\tilde{W}_s|W_s, Z^{mn})-\frac{1}{mn} H(\tilde{W}_s).
\end{align}
Here, \eqref{exp_a} is due to chain rule of conditional entropy, \eqref{exp_b} is due to the definition of mutual information, \eqref{exp_c} comes from the data processing inequality, \eqref{exp_d} is due to the independence of the secret message and the confusion message. In \eqref{exp_e}, we bound each terms on the right hand side as follows:

For the first term, we have $ I(V^{mn};Z^{mn}) \leq  \sum_1^{mn} I(V_i;Z_i)$ $ \leq  mn I(V;Z)$. Therefore, $\frac{1}{mn} I(V^{mn};Z^{mn}) \leq  I(V;Z)$.

To bound the second term, suppose Eve obtains $W_s$ and $Z^{mn}$, and wants to decode $\tilde{W}_s$. By symmetry of chaining construction, Eve can apply similar decoding rule as described in Sec. \ref{Sec_scheme_design}. However, this time Eve decodes from the $m$th block, then the block error probability of the whole \emph{m-chain} block can be upper bounded by
\begin{align}
P_e \leq & (m-1) \sum _{i\in G_{Z \setminus Y}}
               Z(U_i|U^{i-1}, Y^n)   \notag \\
     & +   m    \sum _{i\in  G_{Y \wedge Z} }
               Z(U_i|U^{i-1}, Y^n)
        = O(2^{-n^\beta})
    \notag
\end{align}
for $\beta \in (0, 1/2)$. Hence, by applying Fano's inequality, we have
\begin{align}
H(\tilde{W}_s|W_s, Z^{mn}) &\leq H(P_e) + P_e \log |\tilde{W}_s|  \notag \\
                           & < H(P_e) + P_e[mn I(V;Z)]. \notag
\end{align}
Therefore, as $n \rightarrow \infty $, $\frac{1}{mn} H(\tilde{W}_s|W_s, Z^{mn}) \rightarrow 0$.

For the last term, as $n \rightarrow \infty $, by \eqref{compound_rate} and \eqref{Bob_Eve_rate}, we have $ (m-1)nI(V;Z)  <  H(\tilde{W}_s) < mnI(V;Z) $. Hence, as $m \rightarrow \infty $, $\frac{1}{mn} H(\tilde{W}_s) \rightarrow I(V;Z)$.

After we bound the right hand side of \eqref{exp_e}, we know as $n \rightarrow \infty $ and $m \rightarrow \infty $, $\frac{1}{mn} I(W_s; Z^{mn})\rightarrow 0$. Therefore, the weak secrecy constraint is achieved.

\section{Conclusion}

We proposed a polar coding scheme that achieves the secrecy capacity of the general wiretap channel, by using the chaining construction technique and polar coding for asymmetric channels. Compared to previous work, our construction has better decoding error probability and can be constructed more efficiently. Finally, we note that this chaining construction based polar coding scheme can be extended to achieve \emph{strong} secrecy guarantees as presented in \cite{sasoglu_ISIT_13}.

\vspace*{-0.05cm}

\renewcommand{\baselinestretch}{0.97}

\begin{thebibliography}{10}
\providecommand{\url}[1]{#1}
\csname url@samestyle\endcsname
\providecommand{\newblock}{\relax}
\providecommand{\bibinfo}[2]{#2}
\providecommand{\BIBentrySTDinterwordspacing}{\spaceskip=0pt\relax}
\providecommand{\BIBentryALTinterwordstretchfactor}{4}
\providecommand{\BIBentryALTinterwordspacing}{\spaceskip=\fontdimen2\font plus
\BIBentryALTinterwordstretchfactor\fontdimen3\font minus
  \fontdimen4\font\relax}
\providecommand{\BIBforeignlanguage}[2]{{%
\expandafter\ifx\csname l@#1\endcsname\relax
\typeout{** WARNING: IEEEtran.bst: No hyphenation pattern has been}%
\typeout{** loaded for the language `#1'. Using the pattern for}%
\typeout{** the default language instead.}%
\else
\language=\csname l@#1\endcsname
\fi
#2}}
\providecommand{\BIBdecl}{\relax}
\BIBdecl

\bibitem{Wyner_WTC_75}
A.~D. Wyner, ``The wire-tap channel,'' \emph{Bell System Tech. J.}, vol.~54,
  no.~8, pp. 1355--1387, Oct. 1975.

\bibitem{Csiszar_IT_78}
I.~Csisz{\'a}r and J.~K{\"o}rner, ``Broadcast channels with confidential
  messages,'' \emph{{IEEE} Trans. Inf. Theory}, vol.~24, no.~3, pp. 339--348,
  May 1978.

\bibitem{Arikan_IT_09}
E.~Ar{\i}kan, ``Channel polarization: a method for constructing
  capacity-achieving codes for symmetric binary-input memoryless channels,''
  \emph{{IEEE} Trans. Inf. Theory}, vol.~55, no.~7, pp. 3051--3073, Jul. 2009.

\bibitem{Arikan_ISIT_10}
------, ``Source polarization,'' in \emph{IEEE ISIT}, Jun. 2010.

\bibitem{Korada_IT_10}
S.~Korada and R.~Urbanke, ``Polar codes are optimal for lossy source coding,''
  \emph{{IEEE} Trans. Inf. Theory}, vol.~56, no.~4, pp. 1751--1768, Apr. 2010.

\bibitem{Sasoglu_IT_13}
E.~{\c{S}}a{\c{s}}o{\u{g}}lu, {\.I}.~E. Telatar, and E.~Yeh, ``Polar codes for
  the two-user multiple-access channel,'' \emph{{IEEE} Trans. Inf. Theory},
  vol.~59, no.~10, pp. 6583--6592, Oct. 2013.

\bibitem{Abbe_IT_12}
E.~Abbe and {\.I}.~E. Telatar, ``Polar codes for the {$m$}-user multiple access
  channel,'' \emph{{IEEE} Trans. Inf. Theory}, vol.~58, no.~8, pp. 5437--5448,
  Aug. 2012.

\bibitem{Onay_ISIT_13}
S.~{\"O}nay, ``Successive cancellation decoding of polar codes for the two-user
  binary-input mac,'' in \emph{IEEE ISIT}, Jul. 2013.

\bibitem{Gastpar_arXiv_13}
\BIBentryALTinterwordspacing
N.~Goela, E.~Abbe, and M.~Gastpar, ``Polar codes for broadcast channels,'' Jan.
  2013. [Online]. Available: \url{http://arxiv.org/abs/1301.6150v1}
\BIBentrySTDinterwordspacing

\bibitem{Marco_arXiv_14}
\BIBentryALTinterwordspacing
M.~Mondelli, S.~H. Hassani, I.~Sason, and R.~Urbanke, ``Achieving {M}arton's
  region for broadcast channels using polar codes,'' Feb. 2014. [Online].
  Available: \url{http://arxiv.org/abs/1401.6060v2}
\BIBentrySTDinterwordspacing

\bibitem{Wang_arXiv_14}
\BIBentryALTinterwordspacing
L.~Wang and E.~{\c{S}}a{\c{s}}o{\u{g}}lu, ``Polar coding for interference
  networks,'' Jan. 2014. [Online]. Available:
  \url{http://arxiv.org/abs/1401.7293}
\BIBentrySTDinterwordspacing

\bibitem{Arikan_ISIT_12}
E.~Ar{\i}kan, ``Polar coding for the {S}lepian-{W}olf problem based on monotone
  chain rules,'' in \emph{IEEE ISIT}, Jul. 2012.

\bibitem{Korada_thesis_09}
S.~B. Korada, ``Polar codes for channel and source coding,'' Ph.D.
  dissertation, EPFL, May 2009.

\bibitem{Vardy_IT_11}
H.~Mahdavifar and A.~Vardy, ``Achieving the secrecy capacity of wiretap
  channels using polar codes,'' \emph{{IEEE} Trans. Inf. Theory}, vol.~57,
  no.~10, pp. 6428--6443, Oct. 2011.

\bibitem{Skoglund_Comm_Letter_10}
M.~Andersson, V.~Rathi, R.~Thobaben, J.~Kliewer, and M.~Skoglund, ``Nested
  polar codes for wiretap and relay channels,'' \emph{IEEE Comm. Letters},
  vol.~14, no.~8, pp. 752--754, Aug. 2010.

\bibitem{El_Gamal_PIMRC_10}
O.~O. Koyluoglu and H.~E. Gamal, ``Polar coding for secure transmission and key
  agreement,'' in \emph{IEEE PIMRC}, Sep. 2010.

\bibitem{Shamai_ITW_10}
E.~Hof and S.~Shamai, ``Secrecy-achieving polar-coding,'' in \emph{IEEE ITW},
  Aug. 2010.

\bibitem{Hassani_Allerton_09}
S.~Hassani, S.~Korada, and R.~Urbanke, ``The compound capacity of polar
  codes,'' in \emph{Allerton Conf.}, Sep. 2009.

\bibitem{Vardy_IT_13}
I.~Tal and A.~Vardy, ``How to construct polar codes,'' \emph{{IEEE} Trans. Inf.
  Theory}, vol.~59, no.~10, pp. 6562 -- 6582, Oct. 2013.

\bibitem{Sutter_arXiv_14}
\BIBentryALTinterwordspacing
D.~Sutter and J.~M. Renes, ``Universal polar codes for more capable and less
  noisy channels and sources,'' Apr. 2014. [Online]. Available:
  \url{http://arxiv.org/abs/1312.5990v3}
\BIBentrySTDinterwordspacing

\bibitem{Hassani_arXiv_13}
\BIBentryALTinterwordspacing
S.~H. Hassani and R.~Urbanke, ``Universal polar code,'' Dec. 2013. [Online].
  Available: \url{http://arxiv.org/abs/1307.7223v2}
\BIBentrySTDinterwordspacing

\bibitem{Sasoglu_arXiv_13}
\BIBentryALTinterwordspacing
E.~{\c{S}}a{\c{s}}o{\u{g}}lu and L.~Wang, ``Universal polarization,'' Dec.
  2013. [Online]. Available: \url{http://arxiv.org/abs/1307.7495v2}
\BIBentrySTDinterwordspacing

\bibitem{Honda_IT_13}
J.~Honda and H.~Yamamoto, ``Polar coding without alphabet extension for
  asymmetric models,'' \emph{{IEEE} Trans. Inf. Theory}, vol.~59, no.~12, pp.
  7829--7838, Dec. 2013.

\bibitem{sutter_Arxiv_13}
\BIBentryALTinterwordspacing
D.~Sutter, J.~M. Renes, and R.~Renner, ``Efficient one-way secret-key agreement
  and private channel coding via polarization,'' Apr. 2013. [Online].
  Available: \url{http://arxiv.org/abs/1304.3658}
\BIBentrySTDinterwordspacing

\bibitem{Omur_IT_13}
O.~Ozel and S.~Ulukus, ``Wiretap channels: implications of the more capable
  condition and cyclic shift symmetry,'' \emph{{IEEE} Trans. Inf. Theory},
  vol.~59, no.~4, pp. 2153--2164, Apr. 2013.

\bibitem{Sasoglu_ITW_09}
E.~{\c{S}}a{\c{s}}o{\u{g}}lu and {\.I}.~Telatar, ``Polarization for arbitrary
  discrete memoryless channels,'' in \emph{IEEE ITW}, Oct. 2009.

\bibitem{Mori_ISIT_10}
R.~Mori and T.~Tanaka, ``Channel polarization on {$q$}-ary discrete memoryless
  channels by arbitrary kernel,'' in \emph{IEEE ISIT}, Jun. 2010.

\bibitem{sasoglu_ISIT_12}
E.~{\c{S}}a{\c{s}}o{\u{g}}lu, ``Polar codes for discrete alphabets,'' in
  \emph{IEEE ISIT}, Jul. 2012.

\bibitem{Barg_IT_13}
W.~Park and A.~Barg, ``Polar codes for {$q$}-ary channels, {$q=2^r$},''
  \emph{{IEEE} Trans. Inf. Theory}, vol.~59, no.~2, pp. 955--969, Feb. 2013.

\bibitem{sasoglu_ISIT_13}
E.~{\c{S}}a{\c{s}}o{\u{g}}lu and A.~Vardy, ``A new polar coding scheme for
  strong security on wiretap channels,'' in \emph{IEEE ISIT}, Jul. 2013.

\end{thebibliography}

\end{document}